\documentclass[%
reprint,
 amsmath,amssymb,
 aps,
]{revtex4-2}
\usepackage{graphicx}
\usepackage{dcolumn}
\usepackage{bm}

\usepackage{color}
\usepackage{ulem}
\usepackage{orcidlink}
\usepackage{hyperref}
\hypersetup
{
hypertex=true,
colorlinks=true,
linkcolor=blue,
filecolor=blue,
urlcolor=blue,
citecolor=blue,
breaklinks=true
}
\begin{document}
\preprint{APS/123-QED}
\title{An input-output approach for giant atom scatterings beyond the dipole approximation}
%
\author{S. R. He\orcidlink{https://orcid.org/0000-0002-2837-3410}}
\altaffiliation[]{These authors contributed equally.}

\author{S. N. Wang \orcidlink{https://orcid.org/0000-0002-5498-4047}}
\altaffiliation[]{These authors contributed equally.}

\author{Y. L. Zhang \orcidlink{https://orcid.org/0009-0009-0349-5225}}

\author{P. H. Ouyang\orcidlink{https://orcid.org/0009-0000-5379-8552}}

\author{L. F. Wei
\orcidlink{https://orcid.org/0000-0003-1533-1550}}
\email{lfwei@swjtu.edu.cn}
\affiliation{Information Quantum Technology Laboratory, School of Information Science and Technology, Southwest Jiaotong University, Chengdu 610031, China}

\begin{abstract}
A giant atom is an artificial matter configuration whose spatial scale is comparable to the wavelength of the interacting electromagnetic wave, such that the usual electric-dipole approximation is no longer valid. As a consequence, certain quasi-direct scattering channels for the electromagnetic wave can arise. Given that the well-known input-output approach can only work for the usual point scattering configuration, wherein the electric-dipole approximation is well satisfied, here we develop a modified input-output approach, wherein an additional low-Q cavity channel is introduced, to treat the electromagnetic scattering problem of a giant atom. We demonstrate that, beyond the multiple coupling-point model used widely in recent publications, the present approach can well explain the Fano-type scattering spectra observed generically and extract certain physical parameters, including the energy dissipation parameter of a two-level giant atom and its coupling strength with the scattered electromagnetic wave. Consequently, we argue that various high-performance optical quantum devices, typically the giant-atom-based optical quantum switches, can be generated by engineering the Fano-type scatterings of giant atoms.
\end{abstract}
\maketitle

\section{Introduction}
In the standard theory of quantum optics, the well-known electric-dipole approximation (EDA) is usually adopted to treat various atom-light interactions~\cite{2010OA,2017DR,2017XG,2021AB}, since the wavelength of the radiation is significantly larger than the size of natural atoms, which can therefore generally be treated as point-like quantum emitters. Among various approaches, such as the master equation~\cite{2008DFW,2025SRH}, the real-space approach~\cite{2009JTS,2009JTSS,2024SRH}, and the discrete-coordinate scattering theory~\cite{2008LZ,2011YC}, the input-output approach (IOA) provides one of the most direct ways to analyze the scattering behaviors of atoms~\cite{1984MJC,1985CWG,2015AR,1993CWG,2004CWG,2014KJ,2010SHF,2023ASS,2014MA,2017JC,2013KL,2022SRH,2016ZYL,2019MM}. With this approach, a large number of electromagnetic scattering problems involving cavities~\cite{2021AB,2015AR,1985CWG,2004CWG,2014KJ} and $N$ point-like scatterers, both with and without interactions~\cite{2023ASS,2013KL,2016ZYL,2019MM}, have been successfully investigated.

However, with the continuous development of micro- and nanofabrication technology, a large number of distributed electromagnetic devices~\cite{2008ADO,2025PHO,2025JS}, such as the artificial ``giant atoms''~\cite{2014MVG,2022LD,2014AFK,2016TA,2019GA,2020BK,2021AMV}, have emerged. A basic feature of these devices is that their spatial scales are comparable to the wavelengths of scattered electromagnetic waves, giving rise to nonlocal electromagnetic interactions for which the usual EDA is no longer valid. This provides a new way to manipulate electromagnetic waves. 
Indeed, due to the nonlocal light-matter interactions, distributed electromagnetic devices exhibit a series of novel physical effects that have not been observed in traditional point-like scattering configurations and cannot be explained by the conventional point-like scattering theory.
Specifically, frequency-dependent relaxation rates~\cite{2014AFK,2020BK,2021AMV}, non-exponential decay of excited states~\cite{2019GA}, tunable atom-photon bound states~\cite{2021XW,2022HX,2023AS}, and decoherence-free atom-atom interactions~\cite{2020BK,2018AFK}, among others, have been observed in experiments on electromagnetic scattering involving superconducting giant atoms.

To explain these novel effects beyond the usual EDA, various modified point-like electromagnetic scattering methods have been proposed, including the temporal coupled-mode theory~\cite{2003SHF,2010ZCR,2010QL} and the multiple-scattering-point (MSP) model~\cite{2014MVG,2021QYC,2021FK,2025SQM}. The former is based on the assumption that, in addition to the usual resonant scattering channel, there exists an unknown direct scattering channel. The latter, by contrast, is based on the assumption of a series of hypothetical sequential atom-photon point-like couplings.
However, the scattering spectra predicted by the MSP model still follow the standard Lorentzian line shapes, and therefore fail to explain the asymmetric Fano scattering spectra observed experimentally~\cite{2014MVG,2021AMV}. Although temporal coupled-mode theory can, in principle, be applied to treat these scattering problems beyond the usual EDA and to phenomenologically explain the observed Fano line shapes~\cite{2003SHF,2010ZCR,2010QL}, the scattering parameters associated with such a hypothetical direct scattering channel are still not well defined~\cite{2020HG,2023DR,2021YL}. Therefore, a universal physical model to describe the electromagnetic scattering of giant atoms is still lacking to date.

Given that the usual IOA has been particularly successful in treating point-like scattering within the framework of the usual EDA~\cite{1984MJC,1985CWG,2015AR,1993CWG,2004CWG,2014KJ,2010SHF,2023ASS,2014MA,2017JC,2013KL,2022SRH,2016ZYL,2019MM}, here we propose a possible extension of this approach, referred to simply as the modified input-output approach (MIOA), to explore the electromagnetic scattering of giant atoms beyond the usual EDA. In the proposed MIOA, an additional scattering channel, referred to simply as a “quasi-direct” scattering channel, is introduced alongside the usual point-like resonant scattering channels. Due to the existence of such a “quasi-direct” scattering channel, the scattering spectrum of a giant atom exhibits a Fano line shape that is absent in point-like scattering.
Based on this physical model, we show that the experimentally observed Fano scattering spectra~\cite{2014MVG,2021AMV} of giant atoms can be well fitted. Consequently, the scattering parameters of giant atoms can be accurately extracted. This may provide a basis for the design and application of various Fano devices~\cite{2010AEM,2019DB,2024SSU,2021HL}.

The paper is organized as follows. In Sec.~II, we derive the photon transmission spectrum of the giant-atom scattering model with a quasi-direct scattering channel using a modified input-output approach. We then use this modified model to perform numerical fitting and analysis of the giant-atom scattering spectra reported in Ref.~\cite{2021AMV}. In Sec.~III, we explore two applications of our model, namely, the more precise extraction of the excited-state relaxation time of two-level atoms and high-precision full-band microwave ``on-off'' control. Finally, we summarize our results in Sec.~IV and discuss the potential applications of the proposed method to light-matter interaction systems in which the electric-dipole approximation is invalid.

\section{The MIOA for the Giant atom scatterings}
Let us consider a typical giant-atom configuration shown in Fig.~1, where a bare two-level atom $A$ (with eigenfrequency $\omega_0$) is coupled to scattering photons via $N$ physical coupling points. These actual, rather than hypothetical, coupling points generate the giant-atom configuration $A'$, as depicted in the red dashed box. Since the physical size $D=|x_N-x_1|$ of such a giant atom is comparable to the wavelength $\lambda$ of the scattering photon, the interaction between the giant atom and the scattering photons cannot be treated within the EDA. However, the interaction between the bare atom $A$ and the scattering photons at each of these physical coupling points can still be treated within the usual EDA, while the effects beyond the usual EDA are wholly attributed to the ``quasi-direct'' scattering channel indicated by the green dashed box in Fig.~1. 

Unlike the MSP model with hypothetical multiple coupling points~\cite{2014MVG,2021QYC,2021FK,2025SQM}, the $N$ coupling points considered here represent actual physical coupling points, corresponding to the giant atom configuration demonstrated experimentally in Refs.~\cite{2021AMV,2020BK}, where $N$ takes the values 2, 3, and 6, respectively.
\begin{figure}[ht]
\centering
\includegraphics[width=8cm]{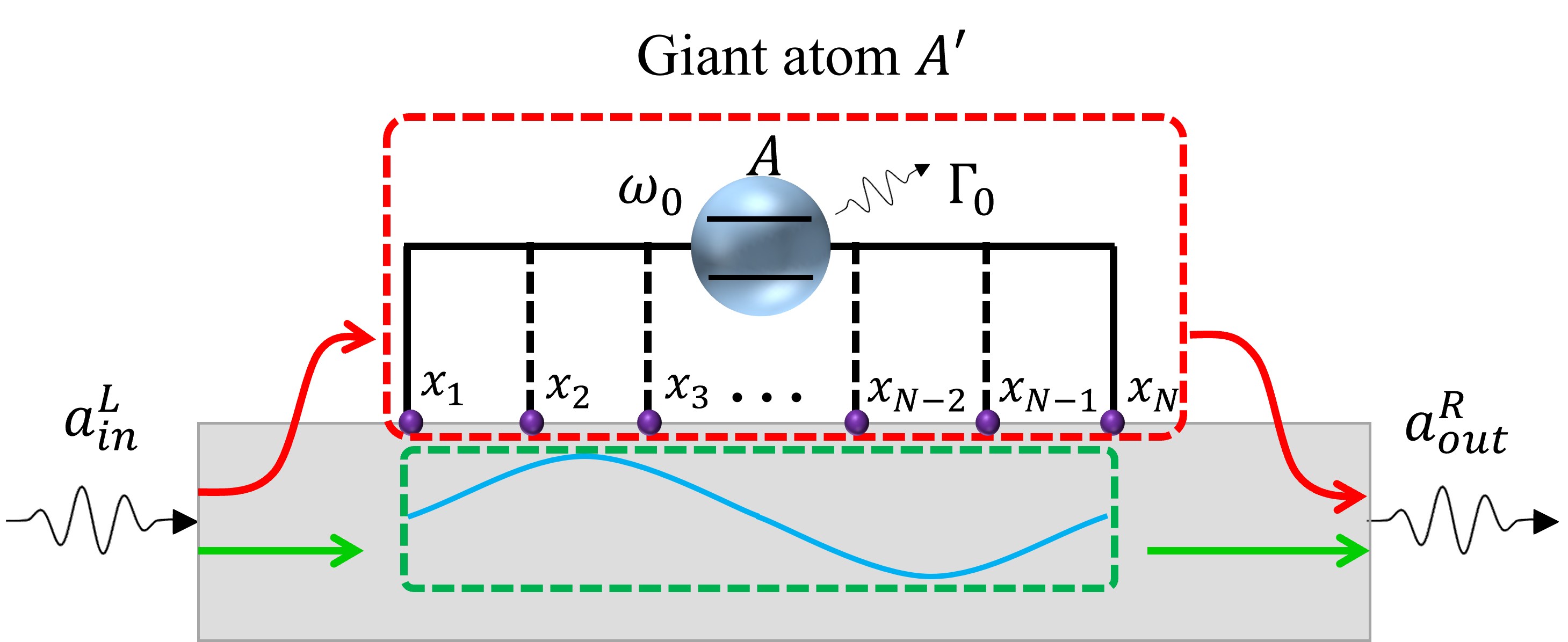}
\caption{A giant atom configuration for the photonic scattering. Here, the red dashed box represents a giant atom configuration formed by a two-level atom, with an eigenfrequency $\omega_0$ and the intrinsic loss $\Gamma_0$, coupled to the scattering photons via $N$ physical (non-virtual) coupling points. The green box indicates the ``quasi-direct'' channel of the scattered photons, where the scattering photons ``bypasses'' the two-level atom.}
\end{figure}
The green dashed box in Fig.~1 is referred to as the “quasi-direct” scattering channel, which describes a process in which a scattering photon can “bypass” the atom $A$ through a quasi-continuous state due to the non-EDA interaction. The corresponding transmission parameter can be expressed as $\mu e^{i\phi}$, with $\mu$ being the transmission amplitude and $\phi$ the transmitted phase shift.
For the usual resonant scattering channel, the scattering photons are coupled to the bare atom $A$ through a series of physical coupling points located at $x_1, x_2, \ldots, x_N$, with $x_1$ and $x_N$ being the input and output points, respectively.

Based on the above analysis, the Hamiltonian for the photonic scattering of a giant atom, depicted in Fig.~1, can be written as $(\hbar=1)$
\begin{equation}
\begin{aligned}
H_{A'}=&(\omega_0-i\frac{\Gamma_0}{2})\sigma_+\sigma_-+\int_{-\infty}^{+\infty}\omega a^\dagger(\omega)a(\omega)d\omega\\
&+i\int_{-\infty}^{+\infty}\sum_{j=1}^Ng_j(\omega)\Big[\sigma_{+}a(\omega)e^{-ikx_j}-H.c.\Big]d\omega\\
&+i\int_{-\infty}^{+\infty}\mu e^{i\phi}\Big[a_{L}(\omega)a_{R}^{\dagger}(\omega)
-H.c.\Big]d\omega.
\end{aligned}
\end{equation}
Here, $\Gamma_0$ denotes the dissipation rate of the bare two-level atom $A$. The second term is the free Hamiltonian of the waveguide photons, where $a^\dagger(\omega)$ and $a(\omega)$ are the corresponding creation and annihilation operators, while $a_L$ and $a_R^\dagger$ denote the annihilation operator for the left-propagating mode and the creation operator for the right-propagating mode, respectively. The third term describes the usual resonant scattering process between the bare atom $A$ and the scattered photons through $N$ actual physical coupling points. Since the spatial extent of each coupling point remains much smaller than the photon wavelength, the local atom–photon interaction at each point can still be treated within the conventional EDA. The phase factor $e^{-ikx_j}$ accounts for the propagation phase associated with the $j$th coupling point, where $k=\omega/v_g$ is the wave vector of the scattered photon, $v_g$ is its group velocity, and $g_j(\omega)$ denotes the atom-photon coupling strength at $x_j$. The last term represents an effective “quasi-direct” scattering channel that originates from nonlocal corrections beyond the EDA. To clarify its physical origin, we consider the light–matter interaction between a giant atom and the waveguide travelling modes:
\begin{equation}
H_I = -\boldsymbol{d} \cdot \big( \boldsymbol{E}_L e^{i k x_l} + \boldsymbol{E}_R e^{i k x_r} \big),
\end{equation}
where $\boldsymbol{E}_L$ and $\boldsymbol{E}_R$ are the electric field amplitudes of the left- and right-going waves, respectively, and $x_l$ and $x_r$ label the two dominant coupling points of the giant atom. For simplicity, we set $\boldsymbol{E}_L=\boldsymbol{E}_R=\boldsymbol{E}_0$ which reduces the interaction to 
$$
H_I = -\boldsymbol{d}\cdot\boldsymbol{E}_0e^{i k x_l} \left[ 1 + e^{i k (x_r - x_l)} \right].
$$
Expanding the relative propagation phase factor yields
\begin{equation}
e^{i k (x_r - x_l)} = 1 + i k (x_r - x_l) - \frac{1}{2} k^2 (x_r - x_l)^2 + \cdots.
\end{equation}
The zeroth-order term recovers the conventional local electric-dipole coupling and is already contained in the third term of $H_{A'}$, while the higher-order terms (involving $x_r - x_l$ and its square) introduce nonlocal corrections. By relating the position fluctuations at the coupling points to the quantized travelling modes, i.e., $x_l \propto (a_L^\dagger + a_L)$ and $x_r \propto (a_R^\dagger + a_R)$, the second-order term $\frac{1}{2} k^2 (x_r - x_l)^2 $ generates two-mode couplings of the form $a_L a_R^\dagger$ and $a_R a_L^\dagger$. Under the rotating-wave approximation, these contributions combine into an effective quasi-direct scattering channel, which is precisely the last term of the Hamiltonian. 
The interference between this “quasi-direct” channel and the usual resonant scattering channel naturally gives rise to the experimentally observed asymmetric Fano line shape. 
The parameters $\mu$
and $\phi$ characterize the transmission amplitude and phase shift of the quasi-direct channel, respectively; a possible microscopic interpretation, along with definitions of $\mu$ and $\phi$ are presented in Appendix~A. Furthermore, in Appendix~B we supplement the description of non-EDA beyond the usual EDA within the distributed resonator scattering model, which is precisely the physical origin of the “quasi-direct” scattering channel (an effect that is completely negligible under the usual EDA). This discussion also serves as an analogy for the origin of the quasi-direct scattering channel in giant atoms.

Obviously, by neglecting the last term in Eq.~(1), the present model can be reduced to the MSP model presented in Refs.~\cite{2014MVG,2021QYC,2021FK,2025SQM}. Furthermore, if we set $d_{j+1,j}=x_{j+1}-x_j\sim 0$, Eq.~(1) describes the usual scattering of a point-like atom, which can be well treated within the usual IOA~\cite{1984MJC,1985CWG,2015AR,1993CWG,2004CWG,2014KJ,2010SHF,2023ASS,2014MA,2017JC,2013KL,2022SRH,2016ZYL,2019MM}. However, due to the existence of the non-EDA effect described by the fourth term in Eq.~(1), the usual IOA for photon scattering must be modified. From the Hamiltonian Eq.~(1) for the scattering of a giant atom configuration shown schematically in Fig.~1, we obtain the Heisenberg equations of motion for the right- and left-propagating photon operators $a_{L/R}$ and the two-level atomic operator $\sigma_{-}$:
\begin{subequations}
\begin{align}
&\frac{d}{dt}a_{R}=-i\omega a_{R}-\sum_{j=1}^Ng_{j}e^{ikx_j}\sigma_{-}+\mu e^{i\phi}a_{L},\\
&\frac{d}{dt}a_{L}=-i\omega a_{L}-\sum_{j=1}^Ng_{j}e^{-ikx_j}\sigma_{-}-\mu e^{i\phi}c_{R},\\
&\frac{d}{dt}\sigma_{-}
=-i(\omega_0-i\Gamma_0)\sigma_{-}-\int_{-\infty}^{+\infty}\sum_{j=1}^Ng_{j}e^{-ikx_j}\sigma_{z}ad\omega.
\end{align}
\end{subequations}

Formally, the solution to Eq.~(4a) can be expressed in the following two forms. For $t_0<t$ (where $t_0$ denotes the initial scattering time), we have:
\begin{equation}
\begin{aligned}
a_{R}(\omega, t>t_{0})
=&a_{R}(\omega, t_{0})e^{-i\omega(t-t_0)}\\
&-\sum_{j=1}^Ng_{j}e^{ikx_j}\int_{t_{0}}^{t}\sigma_-(t')e^{-i\omega(t-t')}dt'\\
&+\mu e^{i\phi}a_{L}(\omega, t_0)\int_{t_{0}}^{t}e^{-i\omega(t-t')}dt',
\end{aligned}
\end{equation}
where $a_{R/L}(\omega, t_{0})$ denotes the right- and left-propagating photon operators at the initial scattering time $t_0$. Similarly, for $t<t_1$ (where $t_1$ denotes the final scattering time), we have:
\begin{equation}
\begin{aligned}
a_{R}(\omega, t<t_{1})
=&a_{R}(\omega, t_{1})e^{-i\omega (t-t_1)}\\
&+\sum_{j=1}^Ng_{j}e^{ikx_j}\int_{t}^{t_1}\sigma_-(t')e^{-i\omega(t-t')}dt'\\
&-\mu e^{i\phi}a_{L}(\omega,t_1)\int_{t}^{t_1}e^{-i\omega (t-t')}dt'.
\end{aligned}
\end{equation}
Similarly, the formal solution to Eq. (4b) can be written as:
\begin{equation}
\begin{aligned}
a_{L}(\omega, t>t_{0})
=&a_{L}(\omega, t_{0})e^{-i\omega(t-t_0)}\\
&-\sum_{j=1}^Ng_{j}e^{-ikx_j}\int_{t_{0}}^{t}\sigma_-(t')e^{-i\omega(t-t')}dt'\\
&-\mu e^{i\phi}a_{R}(\omega, t_0)\int_{t_{0}}^{t}e^{-i\omega(t-t')}dt',
\end{aligned}
\end{equation}
and
\begin{equation}
\begin{aligned}
a_{L}(\omega, t<t_{1})
=&a_{L}(\omega, t_{1})e^{-i\omega(t-t_1)}\\
&+\sum_{j=1}^Ng_{j}e^{-ikx_j}\int_{t}^{t_1}\sigma_-(t')e^{-i\omega(t-t')}dt'\\
&+\mu e^{i\phi}a_{R}(\omega, t_1)\int_{t}^{t_1}e^{-i\omega(t-t')}dt'.
\end{aligned}
\end{equation}

Using the photon operators $a_{L/R}(\omega,t_0)$ of the incident field at the initial time $t_0$ and the photon operators $a_{L/R}(\omega,t_1)$ at the final time $t_1$, the input and output operators of the scattered field are defined as
\begin{equation}
a_{in}^{L/R}(t)=-\frac{1}{\sqrt{2\pi}}\int_{-\infty}^{+\infty} a_{L/R}(\omega,t_0)e^{-i\omega(t-t_0)}d\omega,
\end{equation}
and
\begin{equation}
a_{out}^{L/R}(t)=\frac{1}{\sqrt{2\pi}}\int_{-\infty}^{+\infty} a_{L/R}(\omega,t_1)e^{-i\omega(t-t_1)}d\omega.
\end{equation}
These operators satisfy the commutation relation $[a_{in/out}^{L/R}(\omega), a_{in/out}^{\dagger,L/R}(\omega')]=\delta(\omega-\omega')$. If $g_j$ is assumed to be frequency independent, the effective coupling strength between the giant atom and the scattered photons can be written as $\gamma_j=2\pi g_j^2$. Under the Wigner-Weisskopf approximation, and using Eqs.~(5) and (6), the relation between the input and output fields in the frequency domain for the right-propagating photons is obtained as
\begin{equation}
a_{in}^{R}(\omega)+a_{out}^{R}(\omega)+\mu e^{i\phi}a_{in}^{L}(\omega)=-\sum_{j=1}^N\sqrt{\gamma_{j}}e^{ikx_j}\sigma_-(\omega).
\end{equation}
Similarly, using Eqs.~(7) and (8), the corresponding relation for the left-propagating photons reads
\begin{equation}
a_{in}^{L}(\omega)+a_{out}^{L}(\omega)-\mu e^{i\phi}a_{in}^{R}(\omega)=-\sum_{j=1}^N\sqrt{\gamma_{j}}e^{-ikx_j}\sigma_-(\omega).
\end{equation}

Substituting Eqs.~(5) and (7) into Eq.~(4c), and using the properties of the $\delta$ function,
$\int_{-\infty}^{+\infty} e^{-i\omega(t-t^{\prime})} d\omega/2\pi=\delta(t-t^{\prime})$
and
$\int_{t_0}^t f(t^{\prime})\delta(t-t^{\prime})dt^{\prime}=f(t)/2,$
one obtains the Langevin equation for the operator $\sigma_-$ at the initial time:
\begin{equation}
\begin{aligned}
\frac{d}{dt}\sigma_-=&-i(\omega_0-i\Gamma_0)\sigma_--\sum_{j=1}^N\sqrt{\gamma_{j}}e^{-ikx_j}a_{in}\\
&-\sum_{j=1}^N\frac{\gamma_{j}}{2}e^{-ikx_j}\sum_{j'=1}^Ne^{ikx_{j'}}\sigma_-.
\end{aligned}
\end{equation}
Here, the atom is assumed to be in the weak-excitation regime~\cite{2010SHF}, i.e., $\langle\sigma_z\rangle=-1$. Using the Fourier transform
$x(\omega)=\int_{-\infty}^{+\infty}e^{-i\omega(t-t_0)}x(t)dt/\sqrt{2\pi}$,
with $x(t)=\sigma_-(t),\,a_{in}(t),\,a_{out}(t)$, one obtains the following frequency-domain Langevin equation:
\begin{equation}
\begin{aligned}
&\Big[\sum_{j=1}^N\frac{\gamma_{j}}{2}e^{-ikx_j}\sum_{j'=1}^Ne^{ikx_{j'}}-i(\omega-\omega_0)+\Gamma_0\Big]\sigma_-(\omega)\\
&=-\sum_{j=1}^N\sqrt{\gamma_{j}}e^{-ikx_j}a_{in}(\omega).
\end{aligned}
\end{equation}

Consequently, from Eqs.~(11) and (14), we obtain the scattering amplitude
\begin{equation}
\begin{aligned}
r_{A'}=\frac{\langle a_{out}(\omega) \rangle}{\langle a_{in}(\omega) \rangle}=\frac{i(\omega-\omega_0)+\Gamma_0-\gamma_c/2}{i(\omega-\omega_0)+\Gamma_0+\gamma_c/2}-\mu e^{i\phi},
\end{aligned}
\end{equation}
where
\begin{equation}
\gamma_c=\sum_{j=1}^N\gamma_je^{-ikx_j}\sum_{j'=1}^Ne^{ikx_{j'}}
\end{equation}
is the effective coupling strength for photons scattered by the giant atom.
Similar to the treatment used in the usual point-like scattering model for evaluating the photon transmission probability~\cite{2010SHF,2016ZYL}, we consider the simplest case of a single-port input, where the photons are incident from the left of the coupling point $x_1$. In this case, $a_{in}^R(\omega)=0$, and the transmission probability of the right-propagating photons, injected from the left of $x_1$, scattered by the giant atom, and finally output at $x_N$, can be expressed as
\begin{equation}
T_{A'}(\omega)=\Big|\frac{\langle a_{out}^R(\omega) \rangle}{\langle a_{in}^L(\omega) \rangle}\Big|^2=\Big|\frac{\mathbb{R}(\omega)+1}{2}-\mu e^{i\phi}\Big|^2,
\end{equation}
with
\begin{equation}
\mathbb{R}(\omega)=\frac{i(\omega-\omega_0)+\Gamma_0-\gamma_c/2}{i(\omega-\omega_0)+\Gamma_0+\gamma_c/2}.
\end{equation}
It is straightforward to verify that, if the effect of the ``quasi-direct'' scattering channel is neglected, i.e., $\mu e^{i\phi}=0$, the above result reduces to that of the MSP model in Refs.~\cite{2014MVG,2021QYC,2021FK,2025SQM}. In particular, if the distance between adjacent coupling points is set to $d$ and the atom-photon coupling strength is uniform at each point, i.e., $\gamma_j=\gamma$ and $\theta=kd=\omega d/v_g$, Eq.~(16) can be rewritten as~\cite{2014AFK,2017LG}
\begin{equation}
\tilde{\gamma}_c=\gamma\Big[\frac{\sin(N\theta/2)}{\sin(\theta/2)}\Big]^2.
\end{equation}

Notably, unlike the transmission probability obtained from the MSP model~\cite{2014MVG,2025SQM} based on the usual IOA, the present result contains an additional contribution, $\mu e^{i\phi}$, arising from the ``quasi-direct'' scattering channel. This shows that, once the conventional point-scattering IOA based on the electric-dipole approximation is generalized to giant-atom scattering, an additional non-EDA scattering channel must be included. This constitutes the essential idea of the MIOA proposed here.

Physically, the ``quasi-direct'' channel means that the scattered photons do not simply propagate freely from the input port to the output port, but acquire an additional background scattering amplitude through a quasi-continuous propagation process. As an effective picture, this process can be modeled by a low-$Q$ cavity, for which the scattering parameters defined in Eq.~(17) take the form (see the Appendix~A for details)
\begin{eqnarray}
\left\{
\begin{array}{lll}
\mu(\omega_0,\omega_b)=\frac{\sqrt{\gamma_{b}^L\gamma_{b}^R}}{\sqrt{[(\gamma_{b}^L+\gamma_{b}^R)/2+\Gamma_b]^2+(\omega_0-\omega_b)^2}},\\
\phi(\omega_0,\omega_b)=\tan^{-1}\Big[\frac{(\omega_0-\omega_b)}{(\gamma_{b}^L+\gamma_{b}^R)/2+\Gamma_b}\Big].
\end{array}
\right.
\end{eqnarray}
Here, $\omega_b\sim D^{-1}$ and $\Gamma_0\ll\Gamma_b$, where $\omega_b$, $\Gamma_b$, and $\gamma_b^{L,R}$ are the effective resonance frequency, intrinsic loss, and coupling strengths of the low-$Q$ cavity to the scattered photons at the input and output ports, respectively, and $D$ is the geometric size of the giant atom. Importantly, this description does not require the presence of an actual independent low-$Q$ cavity. Rather, it should be understood as an effective background scattering channel between the left- and right-propagating waveguide modes, which originate from impedance mismatch~\cite{2020HG}, packaging/substrate parasitics\cite{2021WTL}, and other unintended scattering paths\cite{2023DR}

The validity of the giant-atom scattering model based on the above MIOA can be verified by its ability to accurately fit the experimentally observed scattering spectra, which deviate significantly from the standard Lorentzian line shapes.
\begin{figure}[ht]
\centering
\includegraphics[width=8.5cm]{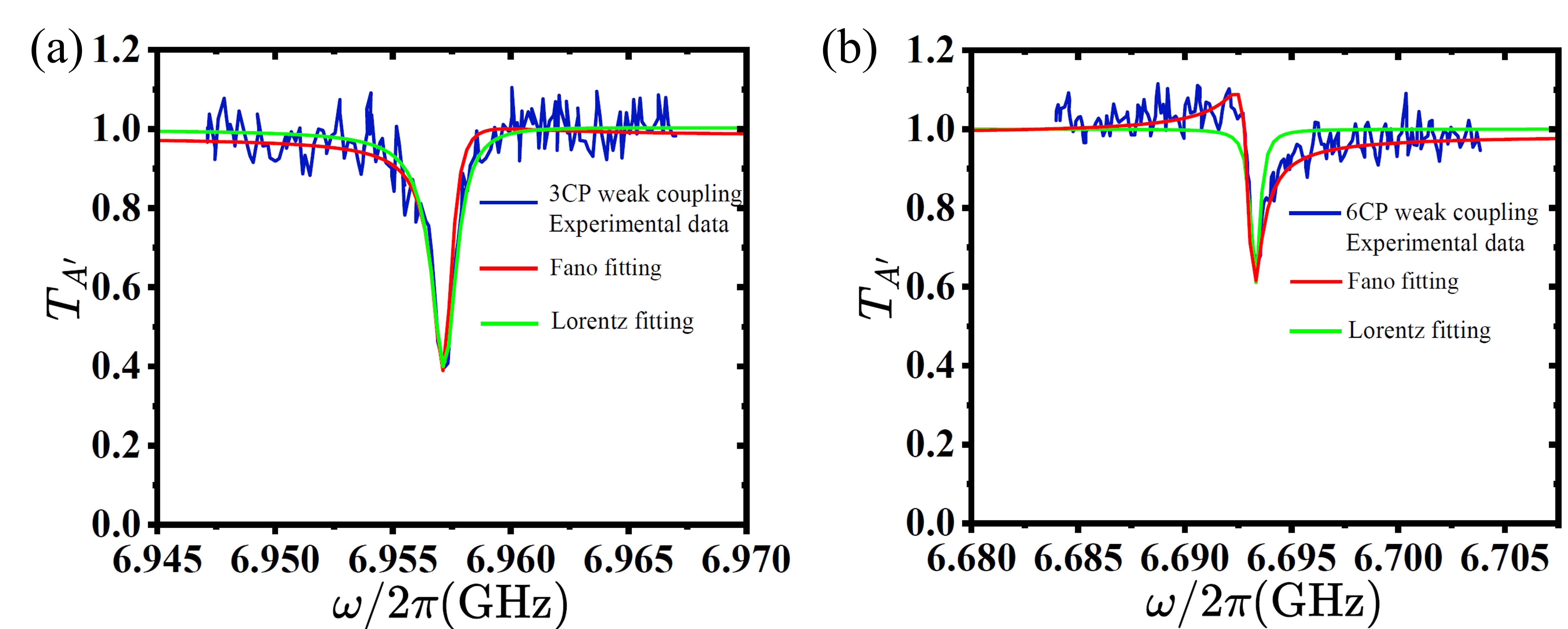}
\caption{Numerical fittings of the experimentally demonstrated spectra of the giant atoms~\cite{2021AMV}: (a) with 3 physical coupling points, and (b) with 6 physical coupling points. Here, the solid blue lines represent the experimentally measured spectra, the green lines denote the fits by using the usual IOA (wherein the ``quasi-direct'' channel scattering effect is neglected), and the red lines stand for the fits by using the MIOA proposed here.}
\end{figure}
Using Eq.~(17), we fit in Fig.~2 the microwave transmission spectra of superconducting qubits reported in Ref.~\cite{2021AMV} for $N=3$ and $N=6$. For comparison, the experimental data, the fits obtained from the MSP model with $\mu e^{i\phi}=0$~\cite{2014MVG,2021QYC}, and those obtained from the present MIOA are shown by the blue, green, and red curves, respectively. Here, the adjacent scattering points are separated by $d=20.54$ mm. The coplanar waveguide resonator is fabricated on a silicon substrate with $\epsilon_{eff}=6.45$, and the microwave group velocity is $v_g=c/\sqrt{\epsilon_{eff}}$, where $c$ is the speed of light~\cite{2021AMV}.
Fig.~2(a), corresponding to $N=3$, uses $\gamma=2\times 10^{-4}\mathrm{GHz}$. The fitting parameters for the observed Fano spectrum are chosen as $\Gamma_0=2.5\times 10^{-4}\mathrm{GHz}$, $\mu=0.429$, and $\phi=-1.03$. By contrast, fitting with a Lorentzian line shape gives $\Gamma_0=3.7\times 10^{-4}\mathrm{GHz}$. Fig.~2(b), corresponding to $N=6$, uses $\gamma=4\times 10^{-6}\mathrm{GHz}$. The fitting parameters in the MIOA are $\Gamma_0=1.2\times 10^{-4}\mathrm{GHz}$, $\mu=0.921$, and $\phi=0.312$, whereas the IOA fit gives $\Gamma_0=2\times 10^{-4}\mathrm{GHz}$.
Evidently, compared with the MSP model under the EDA, the MIOA proposed here provides a better fit to the experimental results. This is because the traditional IOA, which neglects the “quasi-direct” channel, yields only the standard Lorentzian transmission spectrum. To reproduce the experimentally observed non-Lorentzian Fano line shape, the MIOA with an additional adjustable “quasi-direct” term is required~\cite{2003SHF,2010ZCR}.
As explained in the Appendix, this term also admits a reasonable physical interpretation and is not introduced merely as a purely phenomenological fitting term.

\section{Applications of the proposed model}
Since the giant-atom light-scattering model proposed in the present work, based on the MIOA, improves the fit to the experimentally observed scattering spectra, it enables diverse control of light-scattering behavior beyond the usual EDA for various applications.

First, since the scattering spectral formula shown in Eq.~(17) contains the eigenfrequency and dissipation parameters of the two-level atom in the giant atom configuration, the fit to the experimental observations implies that the intrinsic atomic parameters can be extracted more accurately. For instance, using the formula~\cite{2020BK,2024XZ}
\begin{equation}
T_1=(\Gamma_0+\gamma_{c})^{-1},
\end{equation}
the excited-state relaxation time of the two-level atom $A$ in the present giant atom configuration can be determined more precisely by fitting the observed Fano spectra to extract $\Gamma_0$ more accurately, since $\gamma_c$ is fixed by the model. In fact, the extracted value of $\Gamma_0$ differs between the IOA and MIOA fits. Evidently, the value of $\Gamma_0$ obtained from the MIOA fit to the Fano spectra should be closer to the genuine intrinsic loss parameter of the atom.
By fitting the Fano scattering spectra of a giant atom configuration with three coupling points (3CP), reported in Ref.~\cite{2021AMV}, using the MIOA, we obtain $\gamma=0.2$ MHz and $\Gamma_0=0.25$ MHz, and thus $T_1^{3\text{CP}} \approx 2.22~\mu$s. Similarly, for a giant atom configuration with six coupling points (6CP), our fits yield $\gamma=4$ kHz and $\Gamma_0=0.12$ MHz, giving $T_1^{6\text{CP}}\approx 8.06~\mu$s. These values extracted from fitting the observed Fano spectra are consistent with those obtained from the corresponding experimental measurements~\cite{2020BK}.
\begin{figure}[ht]
\centering
\includegraphics[width=6cm]{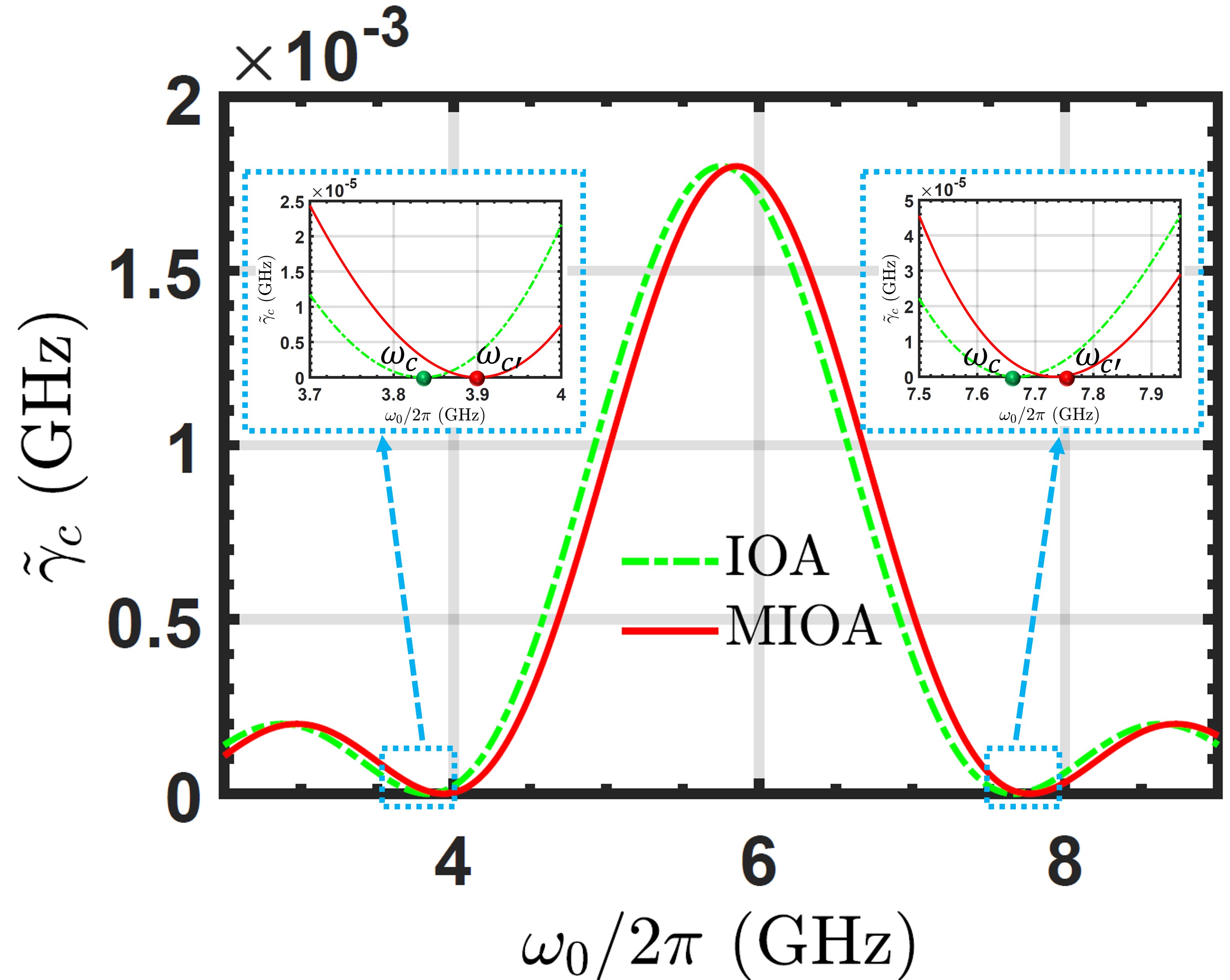}
\caption{The effective coupling strength parameter $\tilde{\gamma}_{c}$, between the scattering photons and the giant atom with 3 physical coupling points~\cite{2021AMV}, for different atomic transition frequency $\omega_0/2\pi$. Here, the green dashed lines and the red solid lines represent the results predicted by the usual IOA and the present MIOA, respectively. The relevant parameters are set as: 
$N=3$, $d=20.45$mm, $\gamma=2\times 10^{-4}\text{GHz}$, $\Gamma_0=2.5\times 10^{-4}\text{GHz}$, $\gamma_{b}^L=0.29\text{GHz}, \gamma_{b}^R=0.31\text{GHz}$, $\Gamma_b=0.01\text{GHz}$, $\omega_b-\omega_0=0.1$GHz, and $\epsilon_{eff}=6.45$.
}
\end{figure}

Secondly, the MIOA can be applied to implement high-precision full frequency-band microwave on-off control by tuning the eigenfrequency $\omega_0$ of the two-level atom. The usual IOA predicts that, if the eigenfrequency $\omega_0$ of the two-level atom is properly chosen to satisfy the condition $\gamma_c(\omega)|_{\omega_0=\omega_c}=0$, the incident electromagnetic wave at any frequency will not be scattered by the giant-atom configuration and can therefore propagate through it freely. However, within the MIOA, this full frequency-band microwave condition must be modified to
\begin{eqnarray}
\big|T_{A'}(\omega)\big|{\omega_0=\omega{c'}}=1,
\end{eqnarray}
where $\omega_{c'}$ denotes the switched-on eigenfrequency of the two-level atom $A$, owing to the contribution from the practically existing “quasi-direct” scattering channel. Specifically, as shown in Fig.~3, using Eq.~(19) together with the parameters determined above, the switched-on eigenfrequencies $\omega_c=3.83,\,7.67$ GHz predicted by the IOA should be corrected to $\omega_{c'}=3.90,\,7.73$ GHz. Evidently, these frequency corrections originate from the non-negligible scattering contribution associated with the additional “quasi-direct” channel.

Basically, beyond the usual EDA, the ``quasi-direct'' scattering effect must be considered. Therefore, the MIOA proposed here should be useful for understanding the physical mechanism of the Fano resonance for the design and application of the various Fano devices.

\section{Conclusions and Discussions}
In summary, we have proposed a MIOA beyond the usual EDA to address various scattering effects of giant atoms. By taking into account the additional “quasi-direct” scattering effect, which is unfortunately neglected in the usual IOA, we find that the experimentally observed Fano resonances can be reproduced more accurately by the proposed MIOA. Compared with temporal coupled-mode theory, we have further introduced a phenomenological low-$Q$ cavity model to provide a clearer physical explanation for the origin of such a “quasi-direct” scattering effect. Therefore, the proposed MIOA is expected to play an important role in the design and application of various Fano devices, including those realized in giant-atom configurations such as the one discussed here.

Although the present work focuses exclusively on the electromagnetic scattering of a single giant-atom system, the MIOA can, in principle, be readily extended to a broader class of light-matter interaction systems in which the usual EDA no longer holds. In particular, it may be applied to multi-giant-atom configurations, as well as to more intricate architectures composed of braided or nested giant atoms.

\section*{Acknowledgments}
This work was partially supported in part by the National Key Research and Development Program of China under Grant No.~2021YFA0718803, the National Natural Science Foundation of China under Grant No.~11974290, the Fundamental Research Funds for the Central Universities under Grant No.~2682024CX048, and the Natural Science Foundation of Sichuan under Grant No.~2025ZNSFSC0857.

\section*{Date availability}
The data are not publicly available. The data are
available from the authors upon reasonable request.

\section*{Appendix A: Derivation of Eq.~(20)}
As shown in Fig.~1, the ``quasi-direct'' scattering of photons can bypass the two-level atom $A$. This implies that the scattered photons can propagate through an additional channel, aside from the scattering via the $N$ physical coupling points. Without loss of generality, such an additional scattering channel beyond the electric-dipole approximation, i.e., the ``quasi-direct'' one (rather than the free propagation), can be modeled as a low-Q cavity, with the resonant frequency $\omega_b$ and the intrinsic loss coefficient $\Gamma_b$. As a consequence, the Hamiltonian (1) could be modified as (with $\hbar=1$):
\renewcommand\theequation{A1}
\begin{equation}
\begin{aligned}
\tilde{H}_{A'}&=(\omega_0-i\frac{\Gamma_0}{2})\sigma_+\sigma_-+\int_{-\infty}^{+\infty}\omega a^\dagger(\omega)a(\omega)d\omega\\
&+i\int_{-\infty}^{+\infty}\sum_{j=1}^Ng_{j}(\omega)\Big[\sigma_{+}a(\omega)e^{-ikx_j}-H.c.\Big]d\omega\\
&+(\omega_b-i\Gamma_b)b^{\dagger}b\\&+i\int_{-\infty}^{+\infty}\sum_{m=L,R}g_{b}^m(\omega)\Big[ba_{m}^{\dagger}(\omega)-H.c.\Big]d\omega,
\end{aligned}
\end{equation}
which replaces the fourth term in Eq.~(1) with the present fourth and fifth terms. Above, $b^\dagger$ and $b$ are the creation and annihilation operators for photons in the low-Q cavity, and $g_{b}^m(\omega)$ ($m=L, R$) represents the coupling strength between the low-Q cavity and the scattered photons at the left and right ports. Clearly, when the influence of the low-Q cavity $b$ is neglected, Hamiltonian (A1) reduces to that for the multiple scattering points model~\cite{2014MVG}. 
From the Hamiltonian (A1), the Heisenberg equations of motion for the operators $a$, $\sigma_{-}$, and $b$ can be written as:
\renewcommand\theequation{A2}
\begin{subequations}
\begin{align}
&\frac{d}{dt}a=-i\omega a-\sum_{j=1}^Ng_{j}e^{ikx_j}\sigma_{-}+\sum_{m=L,R}g_{b}^{m}b,\\
&\frac{d}{dt}b=-i(\omega_b-i\Gamma_b)b-\int_{-\infty}^{+\infty}\sum_{m=L,R}g_{b}^{m}a_md\omega,\\
&\frac{d}{dt}\sigma_{-}
=-i(\omega_0-i\Gamma_0)\sigma_{-}-\int_{-\infty}^{+\infty}\sum_{j=1}^Ng_{j}e^{-ikx_j}\sigma_{z}ad\omega.
\end{align}
\end{subequations}
Following Sec.~II in the main text, the formal solution to Eq.~(A2a) can be expressed as
\renewcommand\theequation{A3}
\begin{equation}
\begin{aligned}
a(\omega,t>t_{0})
&=a(\omega, t_{0})e^{-i\omega(t-t_0)}\\
&-\sum_{j=1}^Ng_{j}e^{ikx_j}\int_{t_{0}}^{t}\sigma_-(t')e^{-i\omega(t-t')}dt'\\
&+\sum_{m=L,R} g_{b}^{m}\int_{t_{0}}^{t}b(t')e^{-i\omega(t-t')}dt',
\end{aligned}
\end{equation}
and
\renewcommand\theequation{A4}
\begin{equation}
\begin{aligned}
a(\omega,t<t_{1})
&=a(\omega, t_{1})e^{-i\omega(t-t_1)}\\
&+\sum_{j=1}^Ng_{j}e^{ikx_j}\int_{t}^{t_1}\sigma_-(t')e^{-i\omega(t-t')}dt'\\
&-\sum_{m=L,R} g_{b}^{m}\int_{t}^{t_1}b(t')e^{-i\omega(t-t')}dt',
\end{aligned}
\end{equation}
respectively. 

In the present model, the scattering of the low-Q cavity and the atom via $N$ physical coupling points are independent. Therefore, by substituting Eq.~(A3) and (A4) into Eq.~(A2b) and (A2c), we obtain the Langevin equations for the operator $\sigma_-$;
\renewcommand\theequation{A5}
\begin{equation}
\begin{aligned}
\frac{d}{dt}\sigma_-=&-i(\omega_0-i\Gamma_0)\sigma_--\sum_{j=1}^N\sqrt{\gamma_{j}}e^{-ikx_j}a_{in}(t)\\
&-\sum_{j=1}^N\frac{\gamma_{j}}{2}e^{-ikx_j}\sum_{j'=1}^Ne^{ikx_{j'}}\sigma_-,
\end{aligned}
\end{equation}
\renewcommand\theequation{A6}
\begin{equation}
\begin{aligned}
\frac{d}{dt}\sigma_-=&-i(\omega_0-i\Gamma_0)\sigma_-+\sum_{j=1}^N\sqrt{\gamma_{j}}e^{-ikx_j}a_{out}(t)\\
&+\sum_{j=1}^N\frac{\gamma_{j}}{2}e^{-ikx_j}\sum_{j'=1}^Ne^{ikx_{j'}}\sigma_-,
\end{aligned}
\end{equation}
and
\renewcommand\theequation{A7}
\begin{equation}
\frac{d}{dt}b=-i(\omega_b-i\Gamma_b)b+\sum_{m=L,R}\sqrt{\gamma_{b}^m}a_{in}^{m}(t)-\sum_{m=L,R}\frac{\gamma_{b}^m}{2}b,
\end{equation}
\renewcommand\theequation{A8}
\begin{equation}
\frac{d}{dt}b=-i(\omega_b-i\Gamma_b)b-\sum_{m=L,R}\sqrt{\gamma_{b}^m}a_{out}^{m}(t)+\sum_{m=L,R}\frac{\gamma_{b}^m}{2}b,
\end{equation}
for the operator $b$.
Where $\gamma_{b}^m=2\pi (g_{b}^m)^2$ represents the effective coupling strength between the photons in the low-Q cavity and the scattered photons at the left and right ports. By making the Fourier transform, we can write the corresponding Langevin equations:
\renewcommand\theequation{A9}
\begin{equation}
\begin{aligned}
&\Big[\sum_{j=1}^N\frac{\gamma_{j}}{2}e^{-ikx_j}\sum_{j'=1}^Ne^{ikx_{j'}}-i(\omega-\omega_0)+\Gamma_0\Big]\sigma_-(\omega)\\
&=-\sum_{j=1}^N\sqrt{\gamma_{j}}e^{-ikx_j}a_{in}(\omega),
\end{aligned}
\end{equation}
\renewcommand\theequation{A10}
\begin{equation}
\begin{aligned}
&\Big[\sum_{j=1}^N\frac{\gamma_{j}}{2}e^{-ikx_j}\sum_{j'=1}^Ne^{ikx_{j'}}-i(\omega_0-\omega)-\Gamma_0\Big]\sigma_-(\omega)\\
&=-\sum_{j=1}^N\sqrt{\gamma_{j}}e^{-ikx_j}a_{out}(\omega),
\end{aligned}
\end{equation}
\renewcommand\theequation{A11}
\begin{equation}
\Big[\sum_{m=L,R}\frac{\gamma_b^m}{2}-i(\omega-\omega_b)+\Gamma_b\Big]b(\omega)=\sum_{m=L,R}\sqrt{\gamma_{b}^m}a_{in}^{m}(\omega),
\end{equation}
and
\renewcommand\theequation{A12}
\begin{equation}
\Big[\sum_{m=L,R}\frac{\gamma_b^m}{2}-i(\omega_b-\omega)-\Gamma_b\Big]b(\omega)=\sum_{m=L,R}\sqrt{\gamma_{b}^m}a_{out}^{m}(\omega),
\end{equation}
in the frequency domain.

By combining Eqs.~(A9)–(A12), the input-output relations for the scattering of the giant atom can be obtained as follows:
\renewcommand\theequation{A13}
\begin{equation}
a_{in}^R(\omega)+a_{out}^R(\omega)=-\sum_{j=1}^N\sqrt{\gamma_{j}}e^{ikx_j}\sigma_-(\omega)+\sqrt{\gamma_{b}^R}b(\omega).
\end{equation}
Specifically, with Eqs.~(A9), (A11), and (A13), we can calculate the scattering probability amplitude:
\renewcommand\theequation{A14}
\begin{equation}
\tilde{r}_{A'}=\frac{i(\omega-\omega_0)+\Gamma_0-\gamma_c/2}{i(\omega-\omega_0)+\Gamma_0+\gamma_c/2}-f(\omega,\omega_b),
\end{equation}
with
\renewcommand\theequation{A15}
\begin{equation}
f(\omega,\omega_b)=\frac{\sqrt{\gamma_{b}^L\gamma_{b}^R}}{
(\gamma_{b}^L+\gamma_{b}^R)/2-i(\omega-\omega_b+i\Gamma_b)},
\end{equation}
for photons injected from the left side.
The second term in Eq.~(A14) arises from the ``quasi-direct'' scattering channel, which is ignored in the usual IOA. Obviously, for the low-Q cavity, $\Gamma_b\gg\Gamma_0$, we expand $f(\omega,\omega_b)$ in a Taylor series, around the atomic transition frequency $\omega_0$, and obtain:
$$
\begin{aligned}
f(\omega_0)\approx&\frac{\sqrt{\gamma_{b}^L\gamma_{b}^R}}{(\gamma_{b}^L+\gamma_{b}^R)/2+\Gamma_b-i(\omega_0-\omega_b)}\\
&+\frac{i\sqrt{\gamma_{b}^L\gamma_{b}^R}}{\Big[(\gamma_{b}^L+\gamma_{b}^R)/2+\Gamma_b-i(\omega_0-\omega_b)\Big]^2}(\omega-\omega_0)
\\&+\cdot\cdot\cdot.
\end{aligned}
$$
Notably, the first-order of this expansion can be neglected for the present low-Q cavity. Thus, we only consider the zeroth-order effect and get
\renewcommand\theequation{A16}
\begin{equation}
f(\omega,\omega_0)\approx \mu(\omega_0,\omega_b) e^{i\phi(\omega_0,\omega_b)},
\end{equation}
where
\renewcommand\theequation{A17}
\begin{eqnarray}
\left\{
\begin{array}{lll}
\mu(\omega_0,\omega_b)=\frac{\sqrt{\gamma_{b}^L\gamma_{b}^R}}{\sqrt{[(\gamma_{b}^L+\gamma_{b}^R)/2+\Gamma_b]^2+(\omega_0-\omega_b)^2}},\\
\phi(\omega_0,\omega_b)=\tan^{-1}\Big[\frac{(\omega_0-\omega_b)}{(\gamma_{b}^L+\gamma_{b}^R)/2+\Gamma_b}\Big].
\end{array}
\right.
\end{eqnarray}
This expression corresponds to Eq.~(20) in the main text. Therefore, modeling the ``quasi-direct'' scattering channel as a low-Q cavity to describe the non-electric-dipole approximation effect is physically reasonable.

\section*{Appendix B: Photon Scattering by a Half-Wavelength Resonator}

As shown in Fig.~4, we consider a half-wavelength resonator with a geometric size $\tilde{D}$ comparable to the wavelength of the electromagnetic field. Because the resonator is no longer a point-like scatterer, the standard EDA is insufficient to fully describe its interaction with the waveguide photons. The effective Hamiltonian can be written as
\renewcommand\theequation{B1}
\begin{equation}
\begin{aligned}
H_c = &(\omega_c - i\Gamma_{\tilde{D}}) c^\dagger c 
+ \sum_{m=L,R} \int_{-\infty}^{+\infty} \omega_m a_m^\dagger(\omega_m) a_m(\omega_m) d\omega_m \\
&+ i \int_{-\infty}^{+\infty} g_{c,n} \Big[ \int_0^{\tilde{D}} \big(c a^\dagger e^{ikx_n} - \text{H.c.}\big) dx_n \Big] d\omega\\
&+i\int_{-\infty}^{+\infty}\tilde{\mu}e^{ik\tilde{\phi}}\Big[a_{L}(\omega)a_{R}^{\dagger}(\omega)
-H.c.\Big]d\omega .
\end{aligned}
\end{equation}
Here, $c^\dagger$ and $c$ are the creation and annihilation operators of the resonator mode with eigenfrequency $\omega_c$ and dissipation rate $\Gamma_{\tilde{D}}$. The parameter $g_{c,n}$ describes the local coupling between the resonator and the waveguide photons at position $x_n$. The last term represents an effective ``quasi-direct'' scattering channel between the left- and right-propagating waveguide modes, with amplitude $\tilde{\mu}$ and phase shift $\tilde{\phi}$.

\begin{figure}[ht]
\centering
\includegraphics[width=8cm]{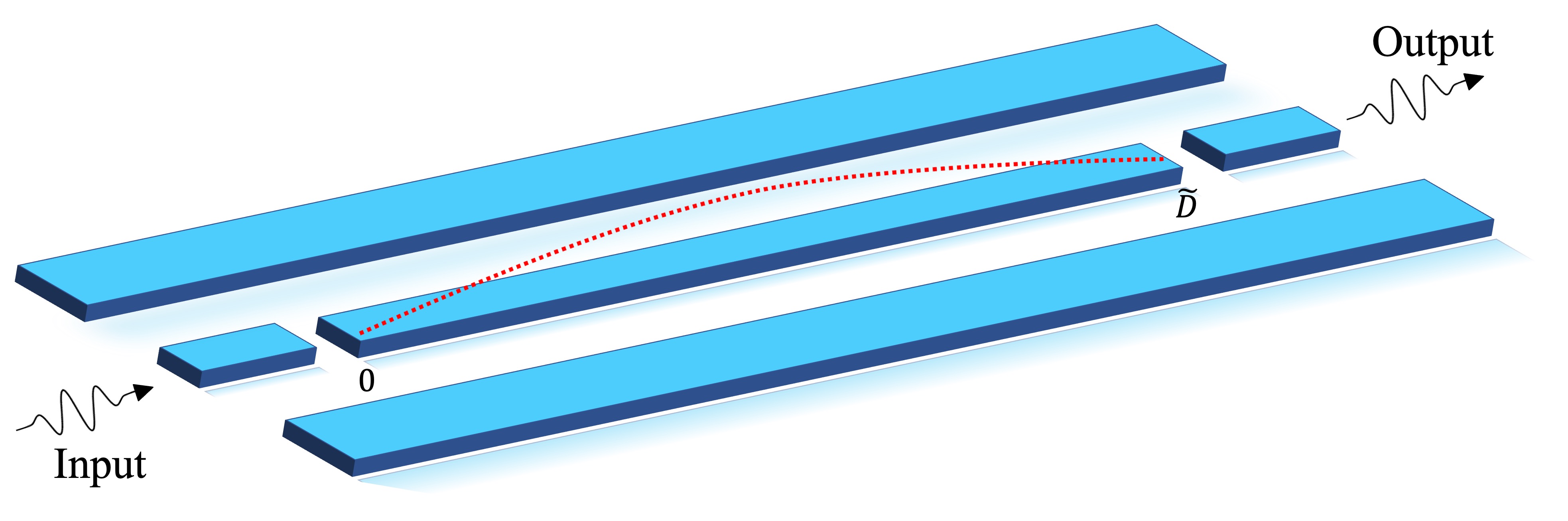}
\caption{Schematic illustration of photon scattering by a half-wavelength transmission-line resonator with a geometric size $\tilde{D}$ comparable to the wavelength of the incident electromagnetic field.}
\end{figure}

The physical origin of this quasi-direct channel can be understood from the nonlocal light-matter interaction between the extended resonator and the travelling waveguide modes. For simplicity, we take the dominant coupling points to be located at $x_l$ and $x_r$. The interaction Hamiltonian is then
\renewcommand\theequation{B2}
\begin{equation}
\tilde{H}_I = -\tilde{\boldsymbol{d}} \cdot \big( \tilde{\boldsymbol{E}}_L e^{i k x_l} + \tilde{\boldsymbol{E}}_R e^{i k x_r} \big),
\end{equation}
where $\tilde{\boldsymbol{E}}_L$ and $\tilde{\boldsymbol{E}}_R$ are the electric-field amplitudes of the left- and right-propagating modes, $\tilde{\boldsymbol{d}}$ denotes the electric dipole moment. Setting $\tilde{\boldsymbol{E}}_L=\tilde{\boldsymbol{E}}_R=\tilde{\boldsymbol{E}}_0$, Eq.~(B2) becomes
\renewcommand\theequation{B3}
\begin{equation}
\tilde{H}_I = -\tilde{\boldsymbol{d}}\cdot\tilde{\boldsymbol{E}}_0 e^{i k x_l}\big[1 + e^{i k (x_r - x_l)}\big].
\end{equation}
Expanding the relative propagation phase gives
$$
e^{i k (x_r - x_l)} =
1 + i k (x_r - x_l)
-\frac{1}{2} k^2 (x_r - x_l)^2 + \cdots .
$$
The zeroth-order term corresponds to the conventional electric-dipole interaction, while the first- and second describe magnetic-dipole, electric-quadrupole contributions, respectively. In the EDA, the scatterer size is much smaller than the wavelength, i.e., 
$\tilde{D}/\lambda \ll 1$, therefore these nonlocal higher-order terms can be neglected and one has
$$
H_{\mathrm{NED}}=-\tilde{\boldsymbol{d}}\cdot\tilde{\boldsymbol{E}}_0\Big[i k (x_r - x_l)
-\frac{1}{2} k^2 (x_r - x_l)^2 + \cdots \Big]\approx 0.
$$
In the present case, however, the resonator size is comparable to the wavelength, $\tilde{D}/\lambda \sim 1$, so the EDA breaks down and $H_{\mathrm{NED}}\neq 0$. These higher-order multipolar corrections provide a microscopic basis for the quasi-direct channel in Eq.~(B1). The connection between the higher-order multipolar corrections and the quasi-direct channel can be further understood. By relating the position fluctuations at the two coupling points to the quantized travelling modes,
$
x_l \propto (a_L^\dagger+a_L),
x_r \propto (a_R^\dagger+a_R),
$
the second-order term in the phase expansion contains
$
(x_r-x_l)^2
\propto
(a_R^\dagger+a_R-a_L^\dagger-a_L)^2 .
$
This expression gives rise to cross-mode coupling terms, including $a_L a_R^\dagger$ and $a_R a_L^\dagger$, together with counter-rotating terms. Under the rotating-wave approximation, only the energy-conserving terms are retained, leading to an effective coupling between the left- and right-propagating modes. Thus, the nonlocal multipolar corrections can be effectively represented by the quasi-direct scattering channel,
which corresponds to the last term of Eq.~(B1).

We next derive the corresponding transmission amplitude using input--output theory. Assuming that photons are incident only from the left port, namely $a_{in}^R=0$, the Heisenberg equation of motion for the resonator operator $c$ is
\renewcommand\theequation{B4}
\begin{equation}
\frac{dc}{dt} = -i(\omega_c - i\Gamma_{\tilde{D}}) c
- \frac{\gamma_{\tilde{D}}}{2} c
+ \sqrt{\frac{\gamma_{\tilde{D}}}{2}} a_{in}^L ,
\end{equation}
where the coupling-induced decay rate is
\renewcommand\theequation{B5}
\begin{equation}
\gamma_{\tilde{D}} =
\Big| \int_0^{\tilde{D}} g_{c,n} e^{ikx_n} dx_n \Big|^2 .
\end{equation}
After Fourier transformation, Eq.~(B4) becomes
\renewcommand\theequation{B6}
\begin{equation}
-i\omega c(\omega) =
-i(\omega_c - i\Gamma_{\tilde{D}}) c(\omega)
- \frac{\gamma_{\tilde{D}}}{2} c(\omega)
+ \sqrt{\frac{\gamma_{\tilde{D}}}{2}} a_{in}^L(\omega) .
\end{equation}
The corresponding input--output relation at the right port is
\renewcommand\theequation{B7}
\begin{equation}
a_{out}^R(\omega) =
\tilde{\mu}e^{ik\tilde{\phi}} a_{in}^L(\omega)
+ \sqrt{\frac{\gamma_{\tilde{D}}}{2}} c(\omega) .
\end{equation}

Combining Eqs.~(B6) and (B7), the transmission amplitude is obtained as
\renewcommand\theequation{B8}
\begin{equation}
t_{\tilde{D}}(\omega)
= \frac{\langle a_{out}^R(\omega) \rangle}
{\langle a_{in}^L(\omega) \rangle}
= \frac{\gamma_{\tilde{D}}/2}
{-i(\Delta + i\Gamma_{\tilde{D}}) + \gamma_{\tilde{D}}/2}
+\tilde{\mu}e^{ik\tilde{\phi}},
\end{equation}
where $\Delta=\omega-\omega_c$ is the detuning between the incident photon and the resonator eigenfrequency. The transmission probability is therefore $T_{\tilde{D}}=|t_{\tilde{D}}(\omega)|^2$.
Eq.~(B8) shows that the total transmission contains two coherent contributions. The first term is the resonant scattering channel dominated by the electric-dipole interaction, while the second term $\tilde{\mu}e^{ik\tilde{\phi}}$, is the quasi-direct background channel generated by higher-order multipolar effects beyond the EDA. These two contributions correspond to the resonant and background components in Eq.~(17) of the main text. Their quantum interference naturally gives rise to the asymmetric Fano line shapes observed in experiments~\cite{2021AB,2008ADO}.

\end{document}